\documentclass[twocolappendix]{emulateapj}

\usepackage[usenames,dvipsnames]{color}
\usepackage{comment}

\shorttitle{GRB Jet Collimation} 
\shortauthors{Duffell, Quataert \& MacFadyen}

\begin{document}

\title{A Narrow Short-Duration GRB Jet from a Wide Central Engine}

\author{Paul C. Duffell\altaffilmark{1}, Eliot Quataert\altaffilmark{1} and Andrew I. MacFadyen\altaffilmark{2}}
\altaffiltext{1}{Astronomy Department and Theoretical Astrophysics Center, University of California, Berkeley, CA 94720}
\altaffiltext{2}{Center for Cosmology and Particle Physics, New York University}
\email{duffell@berkeley.edu}

\begin{abstract}

We use two-dimensional relativistic hydrodynamic numerical calculations to show that highly collimated relativistic jets can be produced in neutron star merger models of short-duration gamma ray bursts without the need for a highly directed engine or a large net magnetic flux.  Even a hydrodynamic engine generating a very wide sustained outflow on small scales can in principle produce a highly collimated relativistic jet, facilitated by a dense surrounding medium which provides a cocoon surrounding the jet core. An oblate geometry to the surrounding gas significantly enhances the collimation process.   Previous numerical simulations have shown that the merger of two neutron stars produces an oblate, expanding cloud of dynamical ejecta.  We show that this gas can efficiently collimate the central engine power much as the surrounding star does in long-duration GRB models.   For typical short-duration GRB central engine parameters, we find jets with opening angles of order 10 degrees in which a large fraction of the total outflow power of the central engine resides in highly relativistic material.  These results predict large differences in the opening angles of outflows from binary neutron star mergers versus neutron star-black hole mergers.

\end{abstract}


\section{Introduction} \label{sec:intro}

Gamma ray bursts (GRBs) involve energetic, collimated relativistic outflows \citep[see][for recent reviews]{2004RvMP...76.1143P, 2006RPPh...69.2259M}.  The current consensus on short-duration GRBs is that they are produced during the merger of a binary neutron star system or a neutron star-black hole system.  However, the detailed properties of the central engine which powers the burst remain uncertain.

GRB engine models often require both fast rotation and large magnetic fields, in order to produce a highly collimated jet.  However, the mechanism for producing a sufficiently strong large-scale field is uncertain, leading to the possibility that it might not be present.  For example, the Blandford-Znajek mechanism typically requires $\sim 10^{15}$ Gauss in large-scale poloidal magnetic field to power a GRB, but such large magnetic flux is not expected to be available in typical neutron star mergers.  It may instead be the case that the magnetic field in the central engine remains toroidal and/or small-scale, and is primarily responsible for providing angular momentum transport in the accretion disk.  From an observational perspective, beaming-corrected GRB energies are inferred to be $\sim 10^{48-50}$ erg \citep{2012ApJ...756..189F, 2014ApJ...780..118F}, yet calculations of typical post-merger disk masses suggest $M_{\rm disk} c^2 \sim 10^{53}$ erg \citep{2013PhRvD..88d4026H}, implying a very inefficient engine without a strong large-scale magnetic flux threading it \citep[contrary to what is often assumed, e.g.][]{Paschalidis:2014qra}.  In fact, in the long GRB context, even if one assumes the existence of strong large-scale poloidal magnetic fields, it can still be difficult for some engine models to produce a highly collimated outflow \citep{2014ApJ...785L..29M}.

These concerns are severe enough that we take seriously the possibility that strong, large-scale magnetic fields may not be present in a short-duration GRB central engine.  In this study, we investigate an alternative picture, in which the engine itself does not need to be strongly collimated, but the outflow becomes a collimated jet after interaction with a dense surrounding medium.  This idea has been previously proposed to explain the formation of radio components in active galactic nuclei \citep{1974MNRAS.169..395B} and is known to be an important part of the collimation process in long GRBs \citep{1999ApJ...524..262M}.

The process of collimation from a wide engine to a narrow jet can be enhanced if the surrounding medium has an oblate geometry \citep[as also suggested by][]{1974MNRAS.169..395B}.  Oblate density profiles are quite natural in GRB scenarios, as most engine models require rapid rotation for operation.

Under the right circumstances, even a spherically symmetric engine can produce a highly collimated outflow if the geometry of the surrounding medium is sufficiently oblate.  This is ultimately due the fact that a blastwave expanding in an asymmetric external medium will preferentially favor the direction of steepest density gradient \citep{1960SPhD....5...46K, 1992ApJ...398..184K, 1999AnA...344..295H}.  Thus, an oblate external medium can result in a prolate blastwave.  If the engine is persistent, continuing to push from the inside of this prolate bubble, the aspect ratio of the bubble will become increasingly exaggerated until the flow is collimated to within a sufficiently small opening angle.  At this point, recollimation shocks can take over \citep{2001ApJ...550..410M, 2003ApJ...586..356Z, 2011ApJ...740..100B} and maintain the collimation until the jet completely breaks out of the surrounding cloud.

In the present study, this collimation process will be demonstrated numerically, in the context of short GRBs.  Numerical relativity simulations show that shortly after a neutron star merger, an oblate cloud of dynamical ejecta is present \citep{2013PhRvD..87b4001H}.  The idea that this cloud can provide a collimating cocoon has also recently been investigated \citep{2014ApJ...784L..28N}, though in that study the parameter space of engines was not thoroughly explored and the surrounding cloud was assumed to be spherical.  Starting with this dynamical ejecta cloud as initial conditions, we investigate the collimation process for a range of engine parameters and including the oblate geometry, and find that very wide hydrodynamic central engines are capable of producing collimated jets.

\section{Numerical Set-Up} \label{sec:numerics}

This study performs a numerical integration of the equations of two-dimensional (2D) axisymmetric relativistic hydrodynamics:

\begin{equation} 
\partial_{\mu} ( \rho u^{\mu} ) = S_D \label{eqn:claw1}
\end{equation} 
\begin{equation} 
\partial_{\mu} ( \rho h u^{\mu} u^{\nu} + P g^{\mu \nu} ) = S^{\nu} \label{eqn:claw2}
\end{equation} 

where $\rho$ is proper density, $\rho h = \rho + \epsilon + P$ is enthalpy density, $P$ is pressure, $\epsilon$ is the internal energy density, $u^{\mu}$ is the
four-velocity, and $g^{\mu \nu}$ is the Minkowski metric (the equations are expressed in units such that $c=1$).  The source terms $S_D$ and $S^{\nu}$ will be used to model the injection of mass, momentum, and energy by the central engine on small scales.  The equation of state is assumed to be radiation dominated: $\epsilon = 3 P$. 

The equations are integrated using the moving-mesh hydrodynamics code, JET \citep{2011ApJS..197...15D, 2013ApJ...775...87D}.  The JET code's moving mesh makes it effectively Lagrangian, which allows for accurate resolution of highly supersonic and relativistic flows.  The inner and outer boundaries are also moved during the flow's evolution, so that the dynamic range covered at any given time is kept modest.  The spatial resolution is given by $\Delta \theta \sim \Delta r / r \sim .006$.  This value is not exact, as zones can expand and contract, but zones are actively refined or de-refined so as to maintain aspect ratios of order unity.  Additionally, zones are more concentrated near the symmetry axis, so that resolution there is about twice as high ($\Delta \theta \sim \Delta r / r \sim .003$).

Initial conditions are modeled after the numerical output of a neutron star merger simulation \citep{2013PhRvD..87b4001H, 2014ApJ...784L..28N}.  It is an outflow described by the following:

\begin{equation} 
\rho( r , \theta , 0 ) = \left\{ \begin{array}
				{l@{\quad \quad}l}
				f(\tilde r) + \rho_{\rm atm}(r) & \tilde r < R_0 	\\  
    			\rho_{\rm atm}(r) & \tilde r > R_0 	\\ 
    			\end{array} \right.    
\end{equation}
\begin{equation} 
v_r( r , \theta , 0 ) = \left\{ \begin{array}
				{l@{\quad \quad}l}
				r/t_0 & \tilde r < R_0 	\\  
    			0 & \tilde r > R_0 	\\ 
    			\end{array} \right.    
\end{equation}
\begin{equation}
f(r) = \rho_c { (1-r/R_0)^{3/4} \over 1 + (r/R_1)^{3.5} } 
\end{equation}
\begin{equation}
\rho_{\rm atm}(r) = \rho_{\rm trans} e^{-r/R_0} + \rho_{\rm ISM}
\end{equation}
\begin{equation}
\tilde r^2 = r^2 ( w^{4/3} cos^2 \theta + w^{-2/3} sin^2 \theta )
\end{equation}

\noindent
where $\rho_c = 287 M_0 / R_0^3$, $R_1 = .06 R_0$, $\rho_{\rm trans} = 10^{-8} M_0/R_0^3$, $\rho_{\rm ISM} = 10^{-20} M_0/R_0^3$, and $t_0 = 10 R_0/c$.  Unless explicitly stated otherwise, the characteristic scales are the initial radius of the cloud $R_0 = 850$ km, and the mass of the cloud $M_0 = 10^{-4} M_{\sun}$, though the hydrodynamical equations are scale-invariant, and therefore the value of these two quantities is arbitrary, so long as all other dimensionful quantities are expressed in terms of these two variables (as well as the speed of light, $c$).  In cgs units, the external density $\rho_{\rm ISM} \approx 3 \times 10^5$ g/cm$^3$ is much larger than typical ISM densities of $\sim 1$ proton per cm$^3$, but this is irrelevant for the current study, as we are only concerned with the production of the jet and not its late-time deceleration.  The density $\rho_{\rm ISM}$ is chosen to be sufficiently low so as to have negligible effect on the jet during the course of the calculation.  The parameter $w = 1.3$ is the aspect ratio of the oblate cloud.

A hydrodynamic engine is injected on small scales, using the source terms $S_D$ and $S^{\nu}$ in equations (\ref{eqn:claw1}) and (\ref{eqn:claw2}), as in \cite{2014arXiv1407.8250D}.  However, in contrast with that study, the engine in the present study will have a very wide half-opening angle, $\theta_0 = 0.5$, or a full opening angle of about 60 degrees.  This is comparable to the opening angle of the high speed outflow seen in hydro simulations of radiatively inefficient black hole accretion disks \citep[e.g.][]{2005ApJ...628..368O} or MHD simulations without significant large-scale magnetic flux \citep[e.g.][]{2014ApJ...796..106J}.  Additionally, in this study, the injection radius $r_0$ is a fixed multiple of the inner boundary, $r_0 = 3 r_{\rm min}$.

The engine duration is set to $\tau = 30 R_0 / c = .085$ seconds.  The engine produces a clean relativistic outflow, with an energy-to-mass ratio of $\eta_0 = 10^{3}$, and a bulk Lorentz factor $\gamma_0 = 10$.  The luminosity of the engine is given by

\begin{equation}
L_0 = e M_0 c^2 / \tau
\end{equation}

\noindent
where $e$ is the ratio of the energy deposited to the cloud rest mass.  Several calculations were performed, varying the value of $e$ from $10^{-3}$ to $0.24$, implying a total energy of $E_{\rm tot} = e M_0 c^2$ in the vicinity of $10^{47}$ - $10^{50}$ ergs (again, the interpreted value of this energy depends on the choice of mass scale, $M_0 = 10^{-4} M_{\sun}$).  The primary calculation assumes $E_{\rm tot} = 4 \times 10^{48}$ erg, and $M_0 = 10^{-4} M_{\sun}$, implying $e = 0.024$.  Short GRB energies have been observationally inferred to be in the $10^{48-50}$ erg range \citep{2012ApJ...756..189F, 2014ApJ...780..118F}.  Alternatively, from theoretical models, typical disk masses are of order $0.1 M_{\sun}$ \citep{2013PhRvD..88d4026H}, so that if this disk is powering the engine, the total energy available to the engine is $\sim 10^{52}$ erg, assuming $\sim 10\%$ efficiency converting the accretion power into an outflow.  The mass of the surrounding cloud is expected to be in the range of $10^{-4} - 10^{-2} M_{\sun}$ \citep{2013PhRvD..87b4001H}, and therefore the value of $e = E_{\rm tot} / M_0 c^2$ could take on values anywhere between $10^{-4}$ and unity, comparable to the range explored in this study.

\begin{figure}
\epsscale{1.2}
\plotone{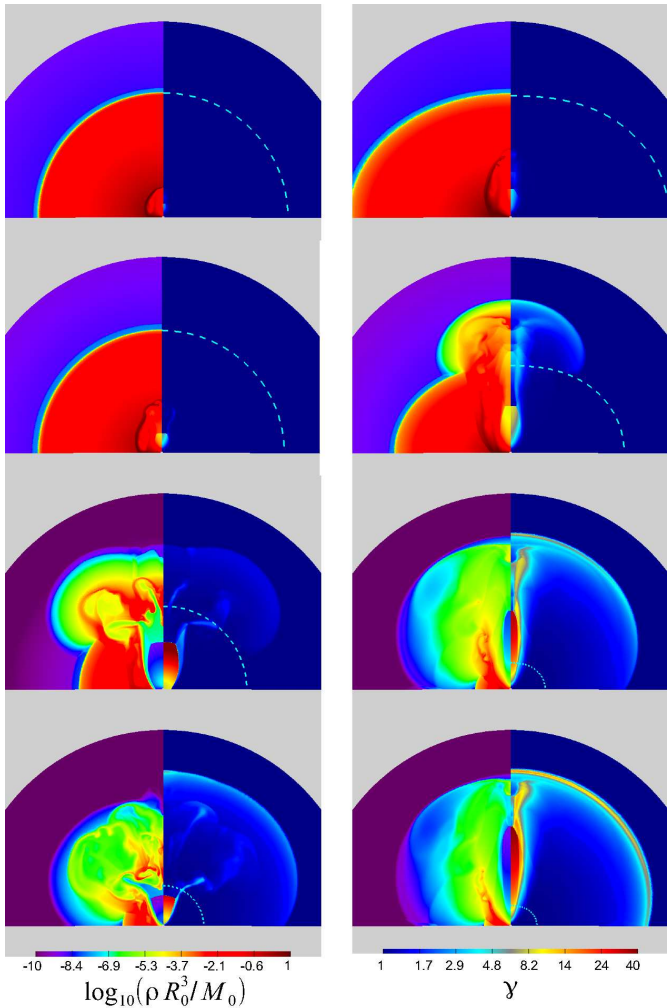}
\caption{ Comparison between oblate and spherical models for the surrounding cloud (spherical on the left side, oblate on the right).  Dashed cyan curves indicate the expanding surface of the cloud of dynamical ejecta.  Various moments in time (top to bottom, $c t / R_0 = 3,5,15,25$) are shown as the engine attempts to clear a passageway in the cloud.  The engine model is injected on a very small scale, which is only barely visible in this figure.  The opening angle of the initial engine is very wide (full opening angle of about 60 degrees), so the spherical cloud is not capable of collimating the flow by itself.  The oblate cloud, however, is sufficient to produce a narrow jet.  In both examples, the ratio of engine energy to ejecta cloud mass is $E_{\rm tot}/M_0 c^2 = 0.024$.
\label{fig:pretty} }
\end{figure}

\begin{figure}
\epsscale{1.2}
\plotone{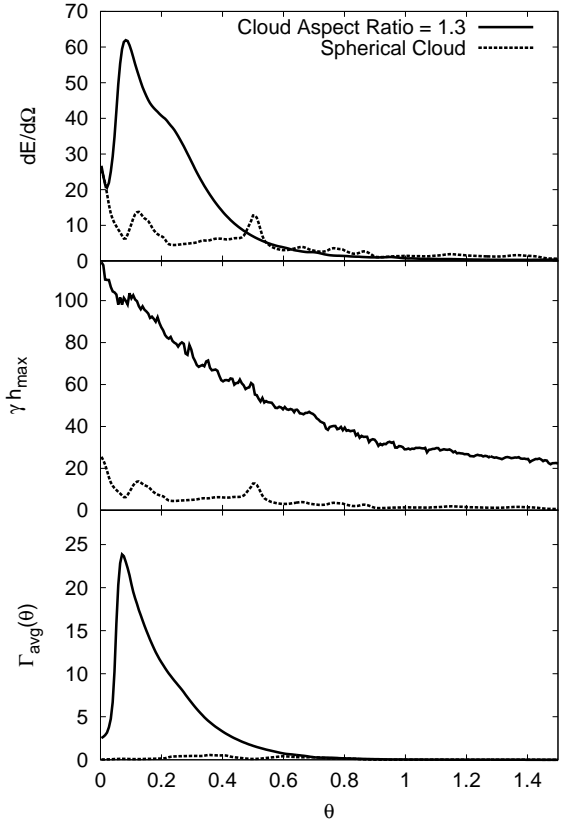}
\caption{ Distribution of energy and terminal Lorentz factor with angle, for both spherical and oblate cloud models from Figure \ref{fig:pretty}.  The spherical model produces a nonrelativistic, spherical outflow, whereas the oblate cloud collimates the jet.  For the oblate model, 50\% of the energy is contained within $\theta < \theta_{50} = 0.12$.  The bottom panel shows an ``averaged Lorentz factor" as a function of polar angle, defined as the ratio of energy to mass at the given angle.  $E_{\rm tot}/M_0 c^2 = 0.024$ in both examples.
\label{fig:theta} }
\end{figure}

\begin{figure}
\epsscale{1.2}
\plotone{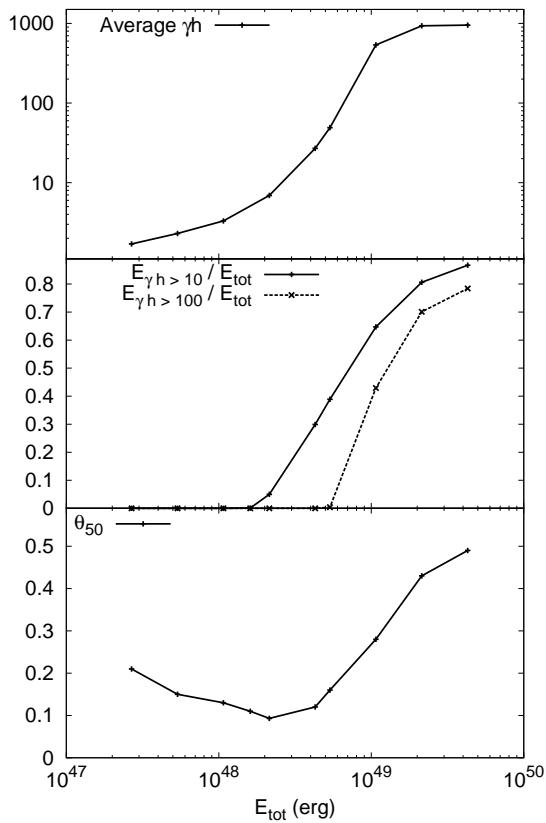}
\caption{ Dependence of terminal Lorentz factor, relativistic energy fractions, and opening angle of the jet on the total energy output of the engine, for a central engine opening angle of $\theta_0 = 0.5$, and an oblate ($w = 1.3$) ejecta cloud mass of $M_0 = 10^{-4} M_{\sun}$.  The relativistic hydro equations are scale-invariant, so these results can be generalized to different choices of $M_0$ by holding $E_{\rm tot}/M_0$ fixed.
\label{fig:edep} }
\end{figure}

\begin{figure}
\epsscale{1.2}
\plotone{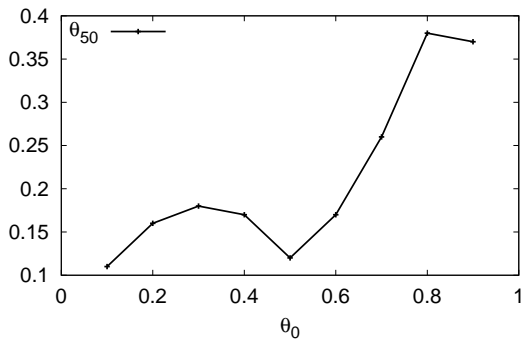}
\caption{ Dependence of the jet half-opening angle, $\theta_{50}$, on the engine half-opening angle, $\theta_0$, injected in the oblate cloud with $w=1.3$ ($e = E_{\rm tot}/M_0 c^2 = 0.024$).  Engines with $\theta_0 \lesssim 0.5$ (corresponding to a full opening angle of 60 degrees) produce collimated jets with $\theta_{50} < 0.2$, roughly independent of $\theta_0$.  Wider engines are more difficult to collimate, though a surrounding cloud which is more oblate ($w > 1.3$) might facilitate collimation of wider engines.
\label{fig:thdep} }
\end{figure}

\begin{figure}
\epsscale{1.2}
\plotone{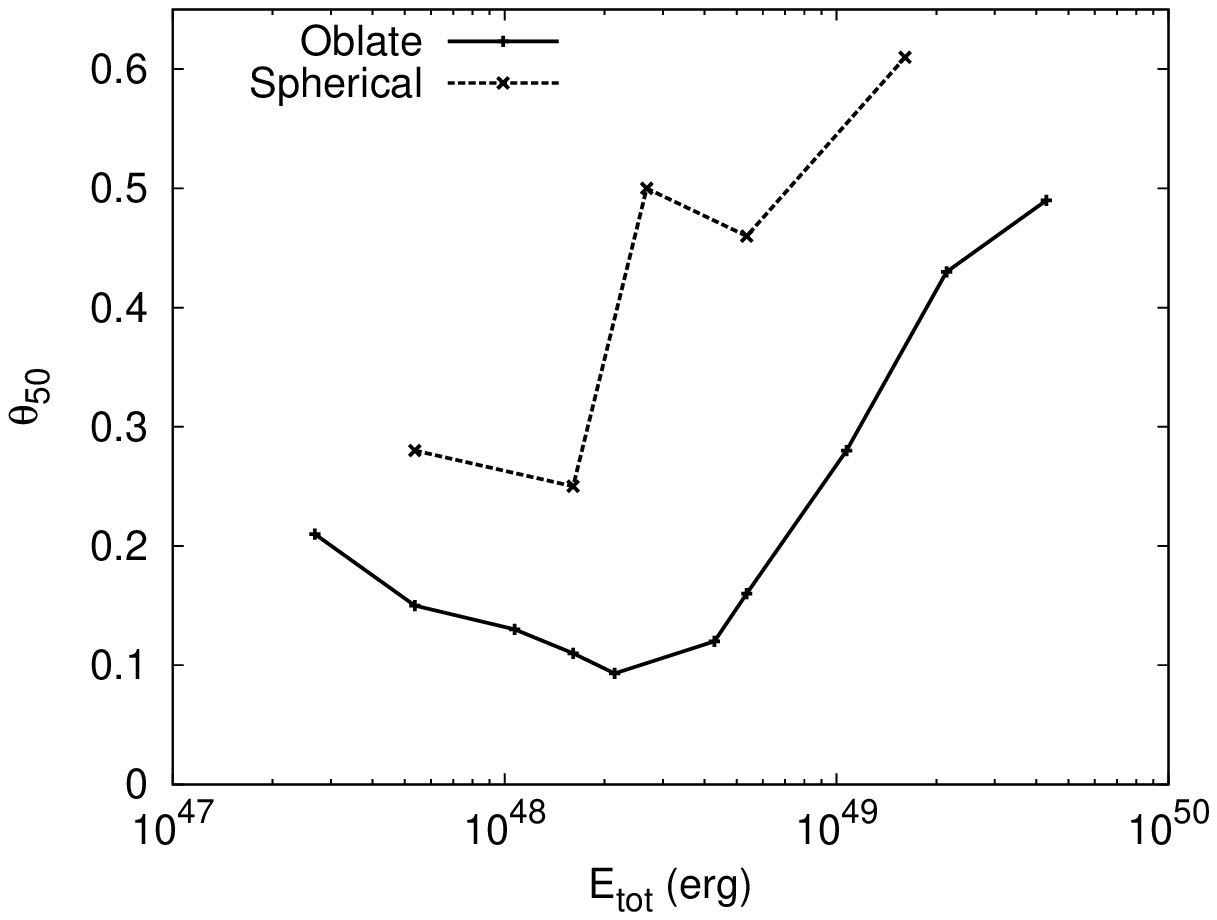}
\caption{ Comparison of the collimation by the spherical cloud ($w=1$) with the oblate cloud ($w=1.3$).  The spherical cloud, at best, can collimate the jet to within $\theta_{\rm 50} \sim 0.25$, but the oblate cloud is more effective at collimation.  This model finds a minimal collimation angle of $\theta_{\rm 50} \sim 0.1$ for the oblate cloud.
\label{fig:w1} }
\end{figure}

\section{Results} \label{sec:results}

Figure \ref{fig:pretty} shows two example runs with $E_{\rm tot} = 2.4 \times 10^{-2} M_0 c^2 = 4.3 \times 10^{48}$ erg.  The engine properties of these two runs are identical; the only difference between the two is the shape of the surrounding cloud, parameterized by the aspect ratio, $w$.  The left side of the figure shows a spherical setup with $w = 1$ and the right side uses $w = 1.3$, a value chosen based on the aspect ratio of the dynamical ejecta found in binary neutron star merger simulations \citep{2013PhRvD..87b4001H}.  The $w = 1$ calculation did not produce a collimated outflow, as the pressure from the cocoon by itself was not significant enough to keep the jet collimated.  The final outflow is roughly spherical (see also Figure \ref{fig:theta}).

In the oblate ($w = 1.3$) run, the outflow from the engine is initially only weakly collimated, but this asymmetry becomes more prominent as the bubble inflates, preferentially moving along the steepest density gradient.  As the engine pushes, this asymmetry becomes increasingly exaggerated until recollimation shocks are capable of keeping the jet collimated.  The flow then clears a passageway, through which a highly collimated jet can escape.  This collimated flow then breaks out of the cloud with a Lorentz factor in the tens, after which post-shock acceleration can increase this Lorentz factor into the hundreds.

The remaining figures show how the flow is distributed in angle and Lorentz factor, and how the results depend on the energy injected by the central engine relative to the mass of the ambient medium (Figures \ref{fig:edep} and \ref{fig:w1}) and the injected opening angle of the outflow (Figure \ref{fig:thdep}).  The terminal Lorentz factor is measured via the specific enthalpy, $\gamma_{\rm terminal} = \gamma h$, which is the terminal Lorentz factor if the fluid element is allowed to expand and adiabatically convert all of its thermal energy to kinetic energy.  In the oblate example shown in Figure \ref{fig:pretty}, about 30\% of the energy has $\gamma h > 10$.

Figure \ref{fig:theta} shows the jet's distribution with latitude.  For the $w = 1.3$ example, $50\%$ of the energy in the jet is contained within $\theta = \theta_{50} = 0.12$.  The $w = 1$ example is also included to show that the flow is roughly spherical and nonrelativistic if the surrounding medium is spherical ($\theta_{50} = 0.46$ for the $w = 1$ model).  Looking back to the $w = 1.3$ case, the dip at very small angles implies a less relativistic jet core (this can also be seen in Figure \ref{fig:pretty}), but this is most likely an artifact of assuming axisymmetry.  The maximum terminal Lorentz factor for this jet is around $\gamma h_{\rm max} \approx 100$.  The ``average Lorentz factor" $\Gamma_{\rm avg}(\theta)$ (bottom panel of Figure \ref{fig:theta}) is calculated via the ratio of total energy to total mass at a given polar angle.

Figure \ref{fig:edep} shows how these results depend on the total energy injected by the engine relative to the mass of the cloud, $E_{\rm tot} / M_0 c^2$.  For sufficiently low energy injection, $E_{\rm tot}$ is negligible compared to the kinetic energy in the dynamical ejecta, and therefore the engine is unsuccessful at producing a collimated outflow.  This leads to the low Lorentz factors and low energy in relativistic material for low values of $E_{\rm tot}$ in Figure \ref{fig:edep}.  For sufficiently large $E_{\rm tot}$, the mass of the ejecta is not significant enough to collimate (or otherwise affect) the flow, and the outflow keeps the same opening angle it was given initially.  Note that the limiting case in the high-energy limit is a jet with $\gamma h \sim 10^3$ and $\theta_{50} \sim 0.5$, roughly the same jet parameters that were injected by the engine, implying that the surrounding cloud does little to affect the properties of the flow.

The behavior in these two limits (low and high energy) imply an intermediate ``sweet spot" for the total energy output of the engine, which is optimal for producing a collimated, relativistic outflow.  For the model presented here, this ``sweet spot" is $e \sim 10^{-2}$, which corresponds to $E_{\rm tot}$ in the few $\times 10^{48}$ erg range for $M_0 = 10^{-4} M_{\sun}$.  This optimal energy likely depends on the detailed structure of gas in the surrounding cloud.  For our engine and cloud model, at this sweet spot, about $10-30\%$ of the energy is used to accelerate the relativistic material with $\gamma h > 10$, while the remaining $70-90\%$ is deposited in the slower-moving material in the dynamical ejecta.

Figure \ref{fig:thdep} shows the dependence of the final opening angle on $\theta_0$, the half-opening angle of the engine (for fixed $E_{\rm tot} / M_0 c^2 = 0.024$).  For sufficiently small $\theta_0$, the outflow has $\theta_{50} \sim 0.1-0.2$, roughly independent of $\theta_0$.  However, above some critical threshold, $\theta_{50}$ begins increasing, so that the final outflow is wide when $\theta_0 \gtrsim 0.7$.

Finally, Figure \ref{fig:w1} compares the spherical ($w = 1$) and oblate ($w = 1.3$) ejecta models, to determine the extent to which oblateness affects collimation.  The spherical ejecta model has difficulty collimating such a wide engine, with a minimum $\theta_{50} \sim 0.25$.  The shape of the surrounding medium therefore plays an important role in the collimation process.

\section{Discussion} \label{sec:disc}

Dynamical processes during neutron star mergers create an expanding oblate cloud of ejecta ($\sim 10^{-4} M_{\sun}$) moving out at $\sim 0.1 c$ \citep{2013PhRvD..87b4001H}.  We have demonstrated that this ejecta collimates outflows from short-duration GRBs, even if the outflow has a large opening angle to begin with.  There is thus no need for a central engine that itself generates a highly collimated relativistic outflow.  This result is a variant of the well-known idea that hot flows can potentially self-collimate as they become supersonic \citep{1974MNRAS.169..395B}.  We also show that the oblateness of the surrounding medium is an important ingredient to the dynamics.

The canonical view of short-duration GRBs has been that they are powered by a relativistic jet analogous to radio-loud quasars or long-duration GRBs \citep{2012ApJ...756..189F, 2014ApJ...780..118F}.  One indication that this analogy breaks down is that the outflow energies in short GRBs appear to be orders of magnitude below the energy available given remnant accretion disk masses of $\sim 0.1 M_{\sun}$ found in simulations of neutron star mergers \citep{2013PhRvD..88d4026H}.  This is in contrast to other jet-dominated sources.  This suggests that the black hole event horizon is not threaded by a dynamically important poloidal magnetic field \citep[i.e. the disk is not in a ``magnetically arrested" state;][]{2011MNRAS.418L..79T}.  Our results are consistent with this general inference in that collimated outflows from short GRBs need not originate from a central engine that in isolation produces a powerful collimated outflow.

Some small-scale and/or toroidal magnetic field is likely still an important part of the engine's operation, but the field does not have to be responsible for collimating the flow to within $\theta_j \sim 0.1$, and the outflow does not have to be magnetically dominated.  A full opening angle of about $60^{\circ}$ appears sufficient for the surrounding medium to collimate the outflow from the central engine (Figures \ref{fig:pretty} and \ref{fig:thdep}).

This result removes some constraints from traditional short GRB engine models, and also opens up new possibilities for what could be powering the central engine.  Neutrino pair annihilation is more viable as an energy source if one does not require the engine itself to be collimated.  A $60^{\circ}$ opening angle is also typical of disk winds, meaning that a disk wind could be a potential engine model, providing that a very clean (baryon-free) wind could be launched from the disk.

The present study focused on the binary neutron star merger scenario for short GRBs.  However, the same considerations apply to neutron star-black hole mergers, for which the dynamical ejecta is expected to be much more oblate \citep[e.g.][]{2014PhRvD..90b4026F}.  Based on the results presented here, we expect differences in the opening angles and Lorentz factors of outflows from black hole-neutron star mergers, relative to binary neutron star mergers.

This result motivates several important future studies.  First, this same calculation should be carried out in full 3D, as it is possible that the collimation and ejection process are sensitive to 3D effects.  For example, the ``hollow jet core" seen in Figure \ref{fig:theta} is likely an artifact of the assumption of axisymmetry.  Second, this calculation should be performed in the context of long GRBs, i.e. a Wolf-Rayet progenitor which is oblate due to rotation \citep[][in preparation]{macduff}.  It seems likely that such a system will exhibit similar behavior.  Third, larger parameter surveys are warranted to study specifically the oblateness of the surrounding medium, and its effect on collimation.  It is also worth noting that Figure \ref{fig:pretty} shows a massive ``plug" or ``cork" in front of the jet, resulting in a ``top-heavy" flow, similar to what was seen in \cite{2014arXiv1407.8250D}.  The properties of this cork and its impact on the jet dynamics should be investigated in more detail.

Finally, it may be possible for analytical jet models \citep{2011ApJ...740..100B, 2014MNRAS.443.1532B} to be modified to include the effect that oblate geometry can have on the properties of a jet breaking out of a surrounding medium.  This would be valuable for understanding what necessary properties an engine must have in order to produce a relativistic, collimated jet after breakout.

\acknowledgments

We thank Francois Foucart for useful conversations.

This work was supported in part by NSF grant AST-1205732, NASA Fermi grant NNX13AO93G, a Simons Investigator award from the Simons Foundation, and the David and Lucile Packard Foundation.  Resources supporting this work were provided by the NASA High-End Computing (HEC) Program through the NASA Advanced Supercomputing (NAS) Division at Ames Research Center.

\bibliographystyle{apj} 

\end{document}